\pdfoutput=1
\documentclass[twocolumn,prd,noshowpacs,superscriptaddress,preprintnumbers,nofootinbib]{revtex4}
\usepackage{graphicx}
\usepackage{epsfig}
\usepackage{bm}
\usepackage{latexsym,amssymb,amsmath,amsfonts,amssymb,txfonts,pxfonts,wasysym,float}
\usepackage{color}

\newcommand{\beq}[1]{\begin{equation}\label{#1}}
\newcommand{\eeq}{\end{equation}}
\newcommand{\bea}[1]{\begin{eqnarray} \label{#1}}
\newcommand{\eea}{\end{eqnarray}}
\newcommand{\ba}{\begin{array}}
\newcommand{\ea}{\end{array}}

\def\be{\begin{equation}}
\def\ee{\end{equation}}
\def\gs{\mathrel{
   \rlap{\raise 0.511ex \hbox{$>$}}{\lower 0.511ex \hbox{$\sim$}}}}
\def\ls{\mathrel{
   \rlap{\raise 0.511ex \hbox{$<$}}{\lower 0.511ex \hbox{$\sim$}}}}

\newcommand{\postscript}[2]{\setlength{\epsfxsize}{#2\hsize}
   \centerline{\epsfbox{#1}}}

\newcommand{\comment}[1]{}

\usepackage[usenames,dvipsnames]{xcolor}
\definecolor{orange}{cmyk}{0,0.5,1,0}
\definecolor{rossoCP3}{cmyk}{0,.88,.77,.40}
\definecolor{graa}{rgb}{0.8,0.8,0.8}
\definecolor{blaa}{rgb}{0.2,0.2,0.6}

\begin{document}

\title{\color{rossoCP3}{Probing strong dynamics with cosmic neutrinos
}}

\author{{\bf Luis A. Anchordoqui}}

\affiliation{Department of Physics and Astronomy,  Lehman College, City University of
  New York, NY 10468, USA
}

\affiliation{Department of Physics,
 Graduate Center, City University
  of New York,  NY 10016, USA
}

\affiliation{Department of Astrophysics,
 American Museum of Natural History, NY
 10024, USA
}

\author{{\bf Carlos Garc\'{\i}a Canal}}
\affiliation{
  Instituto de F\'{\i}sica La Plata, UNLP, CONICET Departamento de F\'{\i}sica, Facultad de Ciencias Exactas, 
Universidad Nacional de La Plata, C.C. 69, (1900) La Plata, Argentina}

\author{{\bf Jorge F. Soriano}}

\affiliation{Department of Physics and Astronomy,  Lehman College, City University of
  New York, NY 10468, USA
}

\begin{abstract}
  \vskip 2mm \noindent IceCube has observed 80 astrophysical neutrino
  candidates in the energy range $0.02 \alt E_\nu/{\rm PeV} \alt 2$.
  Deep inelastic scattering of these neutrinos with nucleons on
  Antarctic ice sheet probe center-of-mass energies
  $\sqrt{s} \sim 1~{\rm TeV}$. By comparing the rates for two classes
  of observable events, any departure from the benchmark (perturbative
  QCD) neutrino-nucleon cross section can be constrained. Using the
  projected sensitivity of South Pole next-generation neutrino
  telescope we show that this facility will provide a unique probe of
  strong interaction dynamics. In particular, we demonstrate that the
  high-energy high-statistics data sample to be recorded by
  IceCube-Gen2 in the very near future will deliver a direct
  measurement of the neutrino-nucleon cross section at
  $\sqrt{s} \sim 1~{\rm TeV}$, with a precision comparable to
  perturbative QCD informed by HERA data.  We also use IceCube data to
  extract the neutrino-nucleon cross section at  $\sqrt{s} \sim 1~{\rm
    TeV}$  through
  a likelihood analysis, considering (for the first time) both the charged-current and
  neutral-current contributions as free parameters of the likelihood function.
\end{abstract}

\maketitle

\section{Introduction}

High-energy neutrinos are unique messengers of far-away phenomena and
can serve as a probe of new physics at sub-fermi distances. {\it Per
  contra} the promise of high energy neutrinos might appear to be
severely limited by astrophysical uncertainties. Event rates constrain
only a combination of fluxes and cross sections, and so astrophysical
uncertainties cloud particle physics implications and vice
versa. However, the event rates for up- and down-going neutrinos
depend differently on neutrino cross
sections~\cite{Kusenko:2001gj,Anchordoqui:2001cg}. By combining both up- and down-going
data one may therefore disentangle particle physics from astrophysics
and constrain both the properties of astrophysical sources and
neutrino interactions.  This technique is entirely agnostic to any
physics process which may modify the neutrino-nucleon cross
section. Essentially this approach constitutes a straightforward
counting experiment.

In this paper we adopt this technique to investigate the sensitivity
of future South Pole neutrino-detection-experiments to the
neutrino-nucleon cross section. Earlier work in this area has
generally assumed a plausible neutrino
luminosity~\cite{Kusenko:2001gj,
  Anchordoqui:2001cg,Anchordoqui:2005pn,Anchordoqui:2006ta,Hussain:2006wg,Borriello:2007cs,Anchordoqui:2018qom}. Now,
however, IceCube measurements yield a non-zero neutrino event rate at
PeV energies~\cite{Aartsen:2013bka, Aartsen:2013jdh,Aartsen:2014gkd,Aartsen:2015zva, Aartsen:2014muf,Aartsen:2017mau}, allowing for a more reliable
calculation.  Indeed, the IceCube Collaboration recently reported a
measurement of the neutrino-nucleon cross section~\cite{Aartsen:2017kpd}. For neutrinos in
the energy bin $6.3 < E_\nu/{\rm TeV} < 980$, the measured cross section is
\begin{equation}
\sigma_{\nu N} = \sigma_{\rm SM}  \times \left[1.30^{+0.21}_{-0.19} {\rm (stat.)} ^{+0.39}_{-0.43}
{\rm (syst.)} \right] \,,
\end{equation}
where $\sigma_{\rm SM}$ is the 
Standard Model (SM) prediction~\cite{CooperSarkar:2007cv,CooperSarkar:2011pa}. 
Further analysis of the IceCube data-sample allowed determination of the energy
dependence of the cross section~\cite{Bustamante:2017xuy}. The proposed
IceCube-Gen2~\cite{Aartsen:2014njl} will surely perform technologically
at least at the level of IceCube, so a conservative estimate of the
sample size is attainable by simply scaling the aperture. IceCube-Gen2
will have an order of magnitude larger aperture than IceCube, which
should provide a sample large enough for a {\it precision} measurement of
the neutrion-nucleon cross section.  Indeed as we show herein
IceCube-Gen2 will be able to determine the neutrino-nucleon cross
section with a precision comparable to perturbative QCD informed by
collider data.

The layout of the paper is as follows. In Sec.~\ref{sec2} we provide
an overview of neutrino detection at IceCube and describe the
different event topologies.  After that we infer the sensitivity of
IceCube to the neutrino-nucleon interaction cross section by combining
upward- and downward-going event rates.  In Sec.~\ref{sec3} we
describe the particulars of our likelihood approach and present the
results from data analysis. We begin by making use of the high-energy reach
of IceCube data to extract the neutrino-nucleon cross section at
energies beyond those available in man-made neutrino beams. As in
previous studies~\cite{Aartsen:2017kpd,Bustamante:2017xuy}, we test
strong dynamics by fixing
 the ratio of charged to neutral current processes to that of the
 perturbative SM. To test non-perturbative SM phenomena, herein we also 
consider the ratio of charged to neutral current processes to be a free
parameter of the likelihood function. Then, armed with our findings,
we investigate the sensitivity of future South Pole
neutrino-detection-experiments to the neutrino-nucleon cross
section. Our conclusions are collected in Sec.~\ref{sec4}.

Before proceeding, it is important to stress that for neutrino energies $\alt 10~{\rm PeV}$, perturbative
QCD provides a robust framework to calculate the neutrino-nucleon
cross section~\cite{Gandhi:1995tf, Gandhi:1998ri,Connolly:2011vc,Chen:2013dza,Arguelles:2015wba, Bertone:2018dse}.  It is only
when the fractional momenta $x$ carried by the constituents become
vanishingly small that the structure functions develop a $\ln(1/x)$
divergent behavior, which in turn results in a violation of unitarity
bounds. Consequently, perturbative QCD predictions are expected to
break down solely when the nucleon has an increasing number of partons with
small $x$.  For the center of mass energies
relevant to our study, however, the neutrino-nucleon cross section can
be calculated perturbatively with an accuracy of better than 5\% when
constrained by measured HERA structure
functions~\cite{CooperSarkar:2007cv,CooperSarkar:2011pa}. Though HERA measurements have
significantly bounded the behavior of neutrino scattering for $E_\nu
\alt 10~{\rm PeV}$, we note that the analysis discussed herein provides
an independent direct measurement of the neutrino-nucleon cross
section in this energy range, and hence is complementary to the
laboratory results.

\section{Neutrino interactions at IceCube}
\label{sec2}

\begin{figure*}
\postscript{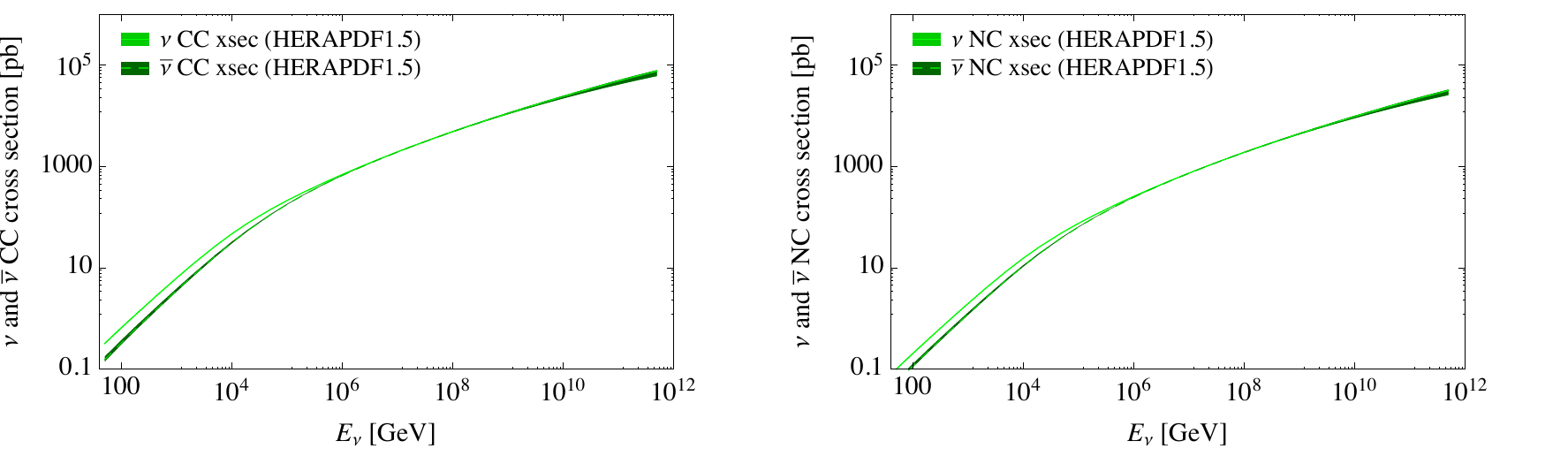}{0.99}
\caption{Neutrino and anti-neutrino cross-sections on isoscalar
  targets for CC and NC scattering according to HERAPDF1.5;
  $\sigma_{{\rm CC},0}$ and $\sigma_{{\rm NC},0}$, respectively. Taken from
Ref.~\cite{CooperSarkar:2007cv}.}
\label{fig:pQCD}
\end{figure*}

Neutrino (antineutrino) interactions in the Antarctic ice sheet can be
reduced to two categories: {\it (i)} in charged current (CC)
interactions the neutrino becomes a charged lepton through the
exchange of a $W^{\pm}$ with some nucleon $N$, $\nu_\alpha (\bar
\nu_\alpha) + N \to \ell_\alpha^\pm + {\rm anything}$; {\it (ii)} in
neutral current (NC) interactions the neutrino interacts via a $Z$
transferring momentum to jets of hadrons, but producing a neutrino
rather than a $\ell^\pm$ in the final state: $\nu_\alpha (\bar
\nu_\alpha)+ N \to \nu_\alpha (\bar \nu_\alpha) + {\rm
  anything}$. Lepton flavor is labeled as $\alpha \in\{e,\mu,\tau\}$
from here on. 

The three neutrino species engender distinctive signal morphologies
when they interact in ice producing the Cherenkov light detected by
the IceCube digital optical modules (DOM), each with a ten-inch
photomultiplier tube (PMT) and associated electronics. The CC
interaction of $\nu_e$ triggers an electromagnetic cascade (or shower)
which ranges out quickly. Such a cascade produces a rather spherically
symmetric signal, and therefore exhibits a low angular resolution of
about $15^\circ - 20^\circ$~\cite{Aartsen:2013jdh}.  However, a fully
or mostly contained shower event provides a relatively precise
measurement of the $\nu_e$ energy, with a resolution of
$\Delta (\log_{10} E_\nu) \approx 0.26$~\cite{Abbasi:2011ui}.  The
situation is reversed for CC interaction $\nu_{\mu}$ induced
events. In this case, the secondary muon travels relatively unscathed
through the ice leaving behind a track.  Muon tracks point nearly in
the direction of the original $\nu_{\mu}$, allowing one to infer the
arrival direction with high angular resolution (say
$\sim 0.7 ^\circ$), while the {\em electromagnetic equivalent energy}
deposited $E_{\rm dep}$ represents only a lower bound of the genuine
$\nu_{\mu}$ energy. For muon tracks, we adopt estimates derived
elsewhere~\cite{Anchordoqui:2016ewn} and set the fractional energy
$E_\mu^{\rm dep}/E_\nu $ to 0.57, 0.51, 0.50, and 0.47 for the IceCube
data set in the interval $10 - 100~{\rm TeV}$, $100-200~{\rm TeV}$,
$200~{\rm TeV}-1~{\rm PeV}$, and $1-10~{\rm PeV}$; respectively. A
point worth noting at this juncture is that the probability
distributions for the parent neutrino energy of a muon track event
which deposits an energy $E_\mu^{\rm dep}$ shown in Fig.~1 of
Ref.~\cite{GonzalezGarcia:2006na} are in good agreement with the
estimates of $E_\mu^{\rm dep}/E_\nu $ adopted herein.  Lastly,
$\nu_{\tau}$ CC interactions may, depending on the neutrino energy,
produce {\it double bang} events~\cite{Learned:1994wg}, with one
shower produced by the initial $\nu_{\tau}$ collision in the ice, and
the second shower resulting from most subsequent $\tau$ decays.
Separation of the two bangs is only feasible for
$E_\nu > 3~{\rm PeV}$, whereas at lower energies the showers tend to
overlap one another. NC interactions of all $\nu$ flavors also produce
showers, but with a smaller rate than CC interactions. For the energy
range of interest, there are two different topologies for the events
registered at IceCube, namely tracks (${\cal T}$) and showers
(${\cal S}$).\footnote{We note in passing that the flavor of a CC
  $\nu_\tau$ interaction of ${\cal S}$ topology (i.e. in which the two
  bangs cannot be separately reconstructed) can be identified by
  searching for {\it double pulse} waveforms that are consistent with
  $\nu_\tau$ CC interaction signatures in IceCube, while rejecting
  waveforms with features that are consistent with late scattered
  photons from single cascade events from NC and $\nu_e$ CC
  interactions~\cite{Aartsen:2015dlt}.} Each of them is produced by
different neutrino flavors and interactions, as summarized in
Table~\ref{tab:topology}.

\begin{table}[t]
\caption{Event topology for each neutrino flavor.}
\begin{tabular}{c|ccc} 
\hline
\hline
~~~Interaction type~~~& ~~~~~~$e$~~~~~~ & ~~~~~~$\mu$~~~~~~ & ~~~~~~$\tau$~~~~~~ \\\hline CC &${\mathcal S}$&${\mathcal
  T}$&${\mathcal S}$\\ NC &${\mathcal S}$&${\mathcal
  S}$&${\mathcal S}$\\ \hline \hline\end{tabular}\label{tab:topology}\end{table}

The classification of observed events in different topologies
  is not always straightforward. While almost all NC $\nu_\mu$ events
  are generally correctly classified as showers, a non negligible
  number of CC $\nu_\mu$ events, of both atmospheric and astrophysical
  origin, could be misclassified as showers if the muon has too
  little energy or is produced near the edge of the detector, escaping
  in both cases without enough energy deposited to be
  detected~\cite{Aartsen:2014muf,Aartsen:2015ivb}.  The effects of
  these misclassifications have been studied in great detail in
  Ref.~\cite{Mena:2014sja,Palomares-Ruiz:2015mka,Vincent:2016nut}.  While
  accounting for misclassifications increases the fraction of
  $\mu$-neutrinos and may have influence on the flavor ratios, with
  present statistics it does not influence neither the shape of the spectrum
  for a shower plus track analysis~\cite{Palomares-Ruiz:2015mka} nor
  cross section studies. In light of this, we
  assume here the event topologies of IceCube high-energy starting events
  (HESE) at face value as given
  in~\cite{Aartsen:2013jdh, Aartsen:2015zva,Aartsen:2017mau}.

The rates at IceCube for down- and up-going events have been
found~\cite{Marfatia:2015hva} to scale respectively as $\Gamma_{\rm down} \propto
\phi \ \sigma_i$ and $\Gamma_{\rm up} \propto \phi \ \sigma_i/\sigma_a$, where
$\phi$ is the neutrino flux, $\sigma_i$ is the cross section for the
interaction that produces the event ($i \in \{{\rm CC}, {\rm NC}\}$), and $\sigma_a$ is the attenuation
cross section, which includes all the effects decreasing the
luminosity due to the fact that neutrinos have to traverse the Earth;
see Appendix~\ref{app1} for details.

For a given bin of energy,  we can constrain neutrino interactions
without assuming particular neutrino fluxes or cross sections. It will
be convenient, however, to present results relative to standard
reference values. IceCube data are consistent with isotropic arrival
directions~\cite{Aartsen:2017ujz} and with expectations for equal
fluxes of all three neutrino
flavors~\cite{Mena:2014sja,Aartsen:2015ivb,Palomares-Ruiz:2015mka,Vincent:2016nut}. For
the reference flux, we adopt the central value of the best-fit power
law of the 4~yr IceCube data~\cite{Aartsen:2014gkd},
\begin{equation}\phi_0(E_\nu)=  2.2 \times 10^{-18}~\left(\frac{E_\nu}{100~{\rm
        TeV}}\right)^{-2.58}~(\mathrm{GeV\,s\,sr\,cm^{2}})^{-1} \, ,
\label{flux0}
\end{equation}
per flavor $\nu_\alpha + \overline \nu_\alpha$. For
the reference cross sections, we choose the results from
perturbative QCD calculations constrained by HERAPDF1.5 shown in
Fig.~\ref{fig:pQCD}. These cross sections have been the benchmarks
adopted by the IceCube Collaboration~\cite{Aartsen:2017kpd}.

For a given flux $\phi$ and cross sections $\sigma_{i}$ and
$\sigma_a$, the expected number of up-going events of a flavor $\alpha$
produced by a charged or neutral current interaction may be expressed as
\begin{subequations}
\begin{equation}N_u^{i,\alpha}\equiv \tilde
  N_u^{i,\alpha}\frac{\phi}{\phi_0}\frac{\sigma_i/\sigma_a^\alpha}{\sigma_{i,0}/\sigma_{a,0}^\alpha},\end{equation}\label{eq:1}
and for down going events,
\begin{equation}
    N_d^{i,\alpha}\equiv \tilde
    N_d^{i,\alpha}\frac{\phi}{\phi_0}\frac{\sigma_i}{\sigma_{i,0}},\label{eq:rates0}\end{equation}
\end{subequations}
with  $i\in\{{\rm
  CC},{\rm NC}\}$ and where the $\tilde N$-constants are obtained assuming that the flux and
cross sections are equal to the reference values, $\sigma_{i,0}$ and $\sigma_{a,0}$.

At this stage it is worthwhile to point out that we have $12N$
quantities ($2 \, {\rm directions} \times 2 \, {\rm interactions}
\times 3 \, {\rm flavors}$), but only 4 of them will be considered in
the data analysis ($2\mbox{ topologies}\times2\mbox{ directions}$). To
gather the events adequately we define the four quantities
\begin{equation} N_x^{\mathcal Z}\equiv \sum_{(i,\alpha)\in {\mathcal
      Z}}N_x^{i,\alpha},\label{eq:2b}\end{equation} with $x\in\{u,d\}$
referring to up- or down-going events, and ${\mathcal Z}\in\{{\mathcal
  T},{\mathcal S}\}$ referring to the event topology (track or shower,
respectively). The sum is extended to the pairs $(i,\alpha)$ contributing
to a topology ${\mathcal Z}$, according to Table~\ref{tab:topology}.

We define $\phi \equiv F\,\phi_0$, $\sigma_{\rm tot} \equiv
S\,\sigma_{{\rm tot},0}$ and the partial cross sections $\sigma_{i,0}
\equiv \alpha_{i,0}\,\sigma_{{\rm tot},0}$ and
$\sigma_i\equiv\alpha_i\,\sigma_{\rm tot}$. The flavor dependent
attenuation cross sections are expressed as $\sigma_{a,0}^\alpha\equiv
a_{\alpha,0}\,\sigma_{{\rm tot},0}$ and $\sigma_a^\alpha\equiv
a_\alpha\,\sigma_{\rm tot}$. The $a$
constants may be expressed in terms of the interaction inelasticities
and the $\alpha$ parameters as $a_{\alpha,0}=\sum_i y_{i,0}^\alpha \alpha_{i,0}$ and
$a_\alpha=\sum_i {y}_i^\alpha \alpha_i$, where $i$ refers to CC or
NC, and $y_i^\alpha$ are the inelasticity parameters for each
interaction~\cite{Gandhi:1995tf}. We can now rewrite (\ref{eq:2b}) as
\begin{subequations}
\begin{equation}
N_d^{\mathcal Z}= F S \sum_{(i,\alpha)\in {\mathcal
    Z}}\frac{\alpha_i}{\alpha_{i,0}} \ \tilde N_d^{i,\alpha},
\end{equation}
\begin{equation} 
N_u^{\mathcal Z}= F  \sum_{(i,\alpha)\in {\mathcal
      Z}}\frac{a_{\alpha,0}  \ \alpha_i}{a_\alpha \ \alpha_{i,0}} \ \tilde N_u^{i,\alpha}.
\end{equation}\label{eq:rates}
\end{subequations}

\begin{figure}
\postscript{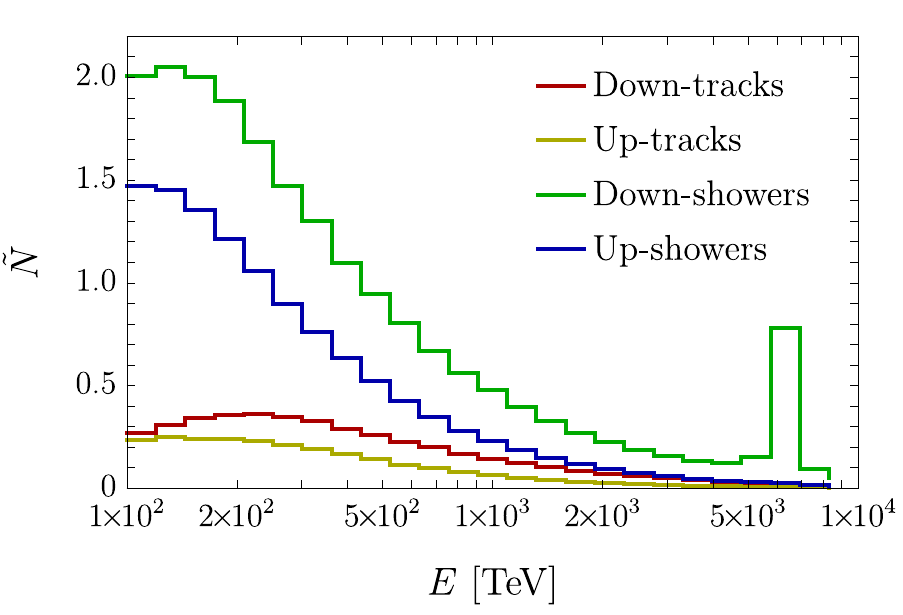}{0.99}
\caption{Reference number of events from (\ref{eq:reference}).}\label{fig:expected}\end{figure}

To perform any further analysis we need to calculate the reference
number of events ($\tilde N_x^{i,\alpha}$) obtained for the flux $\phi_0$ and
cross sections $\sigma_{i,0}$ and $\sigma_{a,0}$ for each of the 12 quantities
involved in (\ref{eq:rates}). This can be done by means of the
expression
\begin{equation}
\tilde N_x^{i,\alpha}\equiv2\pi \,T\int_{E_{\rm
    min}}^{E_{\rm max}} \phi_0(E_\nu) \ A_x^{i,\alpha}(E_\nu)\,\mathrm dE_\nu,
\end{equation}
where $T$ is the running time of the experiment for this sample and
$A_x^{i,\alpha}$ is the effective area averaged for up-(northern) or
down-(southern) going (hemisphere) neutrinos per the flavor $\alpha$
producing an event after a $i$-type interaction. From the IceCube effective
area reported in~\cite{Aartsen:2013jdh}, we obtain the quantity
$A_x^{{\rm CC},\alpha}+A_x^{{\rm NC},\alpha}$. To isolate the
interaction dependence we introduce  the
weights \begin{equation}
w^{i,\alpha}\equiv\frac{\sigma_i
    M_i^\alpha}{\sum_k\sigma_k M_k^\alpha}=\frac{\alpha_i M_i^\alpha}{\sum_k\alpha_k
    M_k^\alpha},\end{equation} where $M_i^\alpha$ is the IceCube target mass for
flavor $\alpha$ and interaction type $i$, given also
in~\cite{Aartsen:2013jdh}. It follows that
\begin{equation}
\tilde
    N_x^{i,\alpha}=w^{i,\alpha}\tilde N_x^\alpha,\end{equation}
and so
\begin{equation}\tilde
    N_x^\alpha\equiv 2\pi\, T\int_{E_{\rm min}}^{E_{\rm max}}
    \phi_0(E_\nu) \ A_x^\alpha(E_\nu)  \ \mathrm
    dE_\nu.
\label{eq:nexpected}
\end{equation}
 The events are distributed in
the same energy bins used in~\cite{Aartsen:2013jdh, Aartsen:2015zva,Aartsen:2017mau}. For the $k$-th
bin, containing events in the energy range $[E_{\rm min}^k,E_{\rm
  max}^k)$, we use in (\ref{eq:nexpected}) the bin averaged effective
area $\left<A_x^\alpha\right>_k$ from~\cite{Aartsen:2013jdh}, and the flux
per flavor given in (\ref{flux0}).
This gives us the reference values in each bin as 
\begin{equation}\tilde N^\alpha_{x,k}\equiv2\pi\,T\left<A_x^\alpha\right>_k\int_{E_{min}^k}^{E_{max}^k}\phi(E_\nu)\,\mathrm d E_\nu.\label{eq:reference}\end{equation}
The values of the expected number of events are shown in
Fig.~\ref{fig:expected}.

In 6 years of observation IceCube has detected above about 30 neutrino
events with energies in the range $0.1 < E_\nu/{\rm PeV} < 2$. This
implies that in 10 years of data taken this facility will collect on
the order of 50 neutrino events within this decade of energy. The next-generation of neutrino telescope in the South pole, IceCube-Gen2, will
increase the per year exposure by about an order of magnitude, and
therefore in 10 yr of observation will collect roughly 500 neutrinos
with $0.1 < E_\nu/{\rm PeV} < 2$.

In the next section we generalize the full-likelihood approach
introduced in~\cite{Anchordoqui:2016ewn} to disentangle cross section
parameters in (\ref{eq:nexpected}) from flux uncertainties in the IceCube data
sample. 

\section{Likelihood analysis}
\label{sec3}

Armed with IceCube observations and expected event rates for fiducial
flux and cross sections we now perform the analysis to extract cross
section parameters using a maximum likelihood method. Let
$\boldsymbol\theta$ be the set of parameters involved in the data
analysis, containing $F$ and all the relevant guidelines to vary the
$\sigma_{{\rm CC},0}$ and $\sigma_{{\rm NC},0}$ cross sections. Let
$\bar N_{x,k}^{\mathcal Z}$ be the measured number of events with
topology ${\mathcal Z}\in\{{\mathcal S},{\mathcal T}\}$ and direction
$x\in\{u,d\}$ in the energy bin $k$. The probability that the bin $k$
contains $\bar N_{x,k}^{\mathcal Z}$ events of type $(x,{\mathcal Z})$
while expecting $N_{x,k}^{\mathcal Z}(\boldsymbol\theta)$ is given by
a Poisson distribution \begin{equation} f\left[\bar N_{x,k}^{\mathcal
      Z}\right.\left|N_{x,k}^{\mathcal
      Z}(\boldsymbol\theta)\right]=\frac{e^{-N_{x,k}^{\mathcal
        Z}}\left(N_{x,k}^{\mathcal Z}\right)^{\bar N_{x,k}^{\mathcal
        Z}}}{\bar N_{x,k}^{\mathcal
      Z}!},\label{eq:poiss}\end{equation} while the probability that
the bin $k$ contains $\bar N_{x,k}^{\mathcal Z}$ events of type
$(x,{\mathcal Z})$ for all the types is \begin{equation}\mathcal
  F_k(\boldsymbol\theta)\equiv\prod_{x,{\mathcal Z}}f\left[\bar
    N_{x,k}^{\mathcal Z}\right.\left|N_{x,k}^{\mathcal
      Z}(\boldsymbol\theta)\right].\end{equation} The likelihood of
having a given a set of parameters $\boldsymbol\theta$ observing the
actual event distribution is \begin{equation}\mathcal
  L(\boldsymbol\theta)=\prod_k\mathcal
  F_k(\theta).\label{eq:like}\end{equation} By the maximization of
$\mathcal L$ in terms of the parameters $\boldsymbol\theta$ we will
estimate the most likely values for those parameters.

We will study several effects that could modify the reference cross
sections.  We parametrize these effects in terms of modifications of
the CC and NC cross sections and their respective inelasticities. Each
particular case would give an expression for $N^{\mathcal
  Z}_{x,k}(\boldsymbol\theta)$ in terms of the reference values
(\ref{eq:reference}) and the parameters $\boldsymbol\theta$. Putting
these expressions in (\ref{eq:like}) will give us the most likely
parameters and the confidence contours in the parameter space. Table~\ref{tab:sm1}
contains the expected number of events in each one of the four
categories compared to the observed ones.

\begin{table}[h]
\caption{Observed/expected number of events in each category.}
\begin{tabular}{c|cc}
\hline
\hline
~~~~~Event direction~~~~~			& ~~~~~~~Shower~~~~~~~	& ~~~~~~~Track~~~~~~~	\\\hline
Down-going	&$18/19.8$		&$6/4.2$		\\\hline
Up-going		&$5/11.5$		&$7/2.5$
\\
\hline
\hline
\end{tabular}\label{tab:sm1}\end{table}

\subsection{Probing strong dynamics with IceCube data}
\label{sec:3a}

The kinematics of lepton-nucleon scattering is described in terms of
the variables $Q^2$, Bjorken $x$, and the inelasticity $y = Q^2/sx$
that measures the energy transfer between the lepton and nucleon
systems, with $s = 2 E_\nu m_N$  the square of the center-of-mass 
energy.  The cross-section for CC neutrino (and antineutrino) scattering
on isoscalar nucleon targets is given
by~\cite{Devenish:2004pb}
\begin{equation}
 \sigma_{{\rm CC},0} =\int_0^1 dx \int_0^{xs} dQ^2 
 {d^2 \sigma^{\nu (\bar \nu) N} \over dx \ dQ^2}\, ,
\label{sigma1}
\end{equation}
where 
\begin{eqnarray}
 {d^2\sigma^{\nu (\bar \nu) N} \over dx \ dQ^2} & = & 
 {G_\mathrm{F}^2 \over 2\pi x}
 \bigg({m^2_W \over Q^2 + m^2_W}\bigg)^2 
 \bigg[Y_+\, F_2^{\nu (\bar \nu)} (x, Q^2) \nonumber \\
 & - & y\, F_{\rm L}^{\nu (\bar \nu)}  (x, Q^2) + Y_-\,
 xF_3^{\nu(\nu)} (x, Q^2)\bigg]
\label{sigma2}
\end{eqnarray}
is the differential cross-section given in terms of the structure
functions $F_2^{\nu (\bar \nu)},$ $F_{\rm L}^{\nu (\bar \nu)}$ and
$xF_3^{\nu (\bar \nu)}$, and $Y_+ = 1 + (1-y)^2$, $Y_- = 1 -
(1-y)^2$. Here, $G_{\rm F}$ is the Fermi constant and $m_W$ is the
$W$-boson mass. At leading order (LO) in perturbative QCD, the
structure functions are given in terms of parton distributions as
$F_2^{\nu (\bar \nu)}
= x [\sum_i \alpha_i q_i(x,Q^2) + \sum_j \alpha_j \bar{q}_j(x, Q^2)],$
$xF_3^{\nu (\bar \nu)} =
x[ \sum_i  \beta_i q_i(x, Q^2) + \sum_j \beta_j \bar{q}_j(x, Q^2)]$
and $F_{\rm L}^{\nu (\bar \nu)} =
0$~\cite{Devenish:2004pb}. For neutrinos, $i = u,d,s,b$ and $j= u,d,c$, 
 with $\alpha_i = \alpha_j = \beta_i  = 1$ for $u,d$; $\alpha_i=\alpha_j = \beta_i = 2$ for
 $s,b$; $\beta_j = -1$ for $u,d$; $\beta_j = -2$ for $c$ quarks.  For antineutrinos,
$i = u,d,c$ and $j = u, d,s,b$, with $\alpha_i = \alpha_j = \beta_i  = 1$ for $u,d$; $\alpha_i=\alpha_j = \beta_i = 2$ for
 $c$; $\beta_j = -1$ for $u,d$; $\beta_j = -2$ for $s,b$ quarks.

 The NC cross sections on isoscalar targets are given by expressions
 similar to (\ref{sigma1}) and (\ref{sigma2}), with the $W$ propagator
 replaced by the $Z$ propagator. For NC interactions the LO
 expressions for the structure functions are given by $F_2^{\nu (\bar
   \nu)} = x \{ \sum_i \alpha_i [q_i(x,Q^2) + \bar{q}_i(x, Q^2)] +
 \sum_j \alpha_j [q_j(x,Q^2) + \bar{q}_j(x, Q^2) ]+ \sum_k \alpha_k[
 q_k(x,Q^2) + \bar{q}_k(x, Q^2)] \}$ and $xF_3^{\nu (\bar \nu)} =
 \sum_i x (v_ua_u + v_d a_d) [ q_i(x, Q^2) - \bar{q}_i(x, Q^2)]$,
 where $i = u,d$, $j= s,b$, $k = c$, $\alpha_i = (a_u^2 +v_u^2 +a_d^2
 +v_d^2)/2$, $\alpha_j = a_d^2 +v_d^2$, and $\alpha_k =a_u^2 +v_u^2$,
 with $v_u$, $v_d$, $a_u$, $a_d$ the NC vector and axial couplings for
 $u-$ and $d$-type quarks~\cite{Devenish:2004pb}.

\begin{figure}[tpb]
\postscript{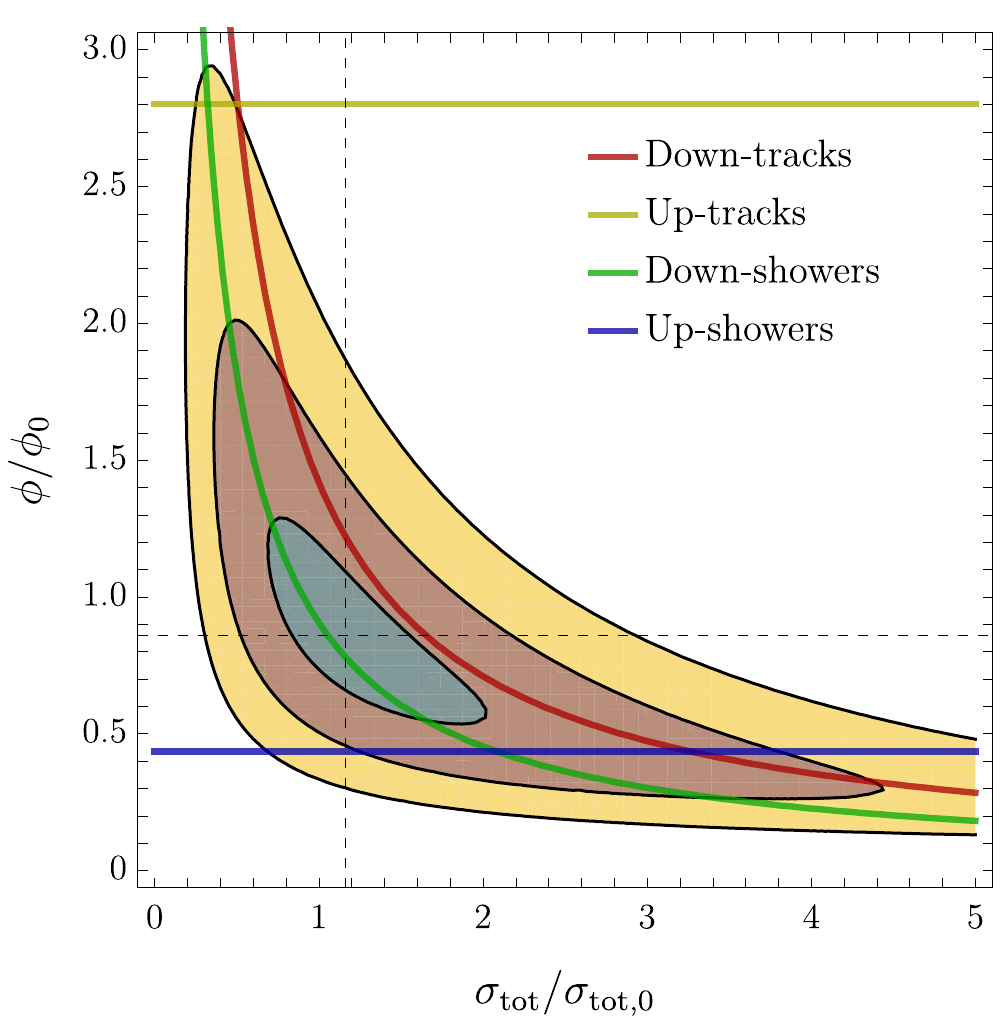}{0.9}
\caption{1, 3, and $5\sigma$ confidence contours for $(F,S)$ for
  scaled total cross section $\sigma_{\rm tot}$ and flux $\phi$ with
  respect to their reference values $\sigma_{{\rm tot},0}$, $\phi_0$.}\label{fig:scale}\end{figure}

At next-to-leading order (NLO) the $F$-functional relations involve
further QCD-calculable coefficient functions and contributions from
$F_{\rm L}$ can no longer be neglected~\cite{Arguelles:2015wba}. The
parton distribution functions (PDFs) are determined in fits to deep
inelastic scattering (DIS) data by the following procedure. The PDFs
are parameterized at some initial scale $Q_0 \sim 1$~GeV and then
evolved, using the NLO DGLAP equations~\cite{Altarelli:1977zs,Gribov:1972ri,Lipatov:1974qm,Dokshitzer:1977sg}, to
higher values of $Q^2$. They are then convoluted with QCD-calculable
coefficient functions to give NLO predictions for the structure
functions, which are then fitted to the DIS data, to obtain the CC and
NC neutrino-nucleon cross sections shown in
Fig.~\ref{fig:pQCD}~\cite{CooperSarkar:2007cv}.

To probe the PDFs,  we assume a simple global scaling of the total reference cross section,
$\sigma_{\rm tot} =S \sigma_{{\rm tot},0}$, and thus
$\alpha_i=\alpha_{i,0}$. We further assume the inelasticity 
of the NC interaction remains unchanged, and so
$a_{\alpha_0}=a_\alpha$. With this in mind, the
set of parameters for the likelihood analysis is
$\boldsymbol\theta=\{F,S\}$, and  the expressions in
(\ref{eq:rates}) become
\begin{subequations}
\begin{equation}N_d^{\mathcal Z}=FS\sum_{(i,\alpha)\in {\mathcal Z}}\tilde N_d^{i,\alpha},\end{equation}
\begin{equation} N_u^{\mathcal Z}=F  \sum_{(i,\alpha)\in {\mathcal
      Z}}\tilde N_u^{i,\alpha},\end{equation}\end{subequations} for
$\mathcal Z\in\{\mathcal S,\mathcal T\}$.

\begin{figure}[tpb]
\postscript{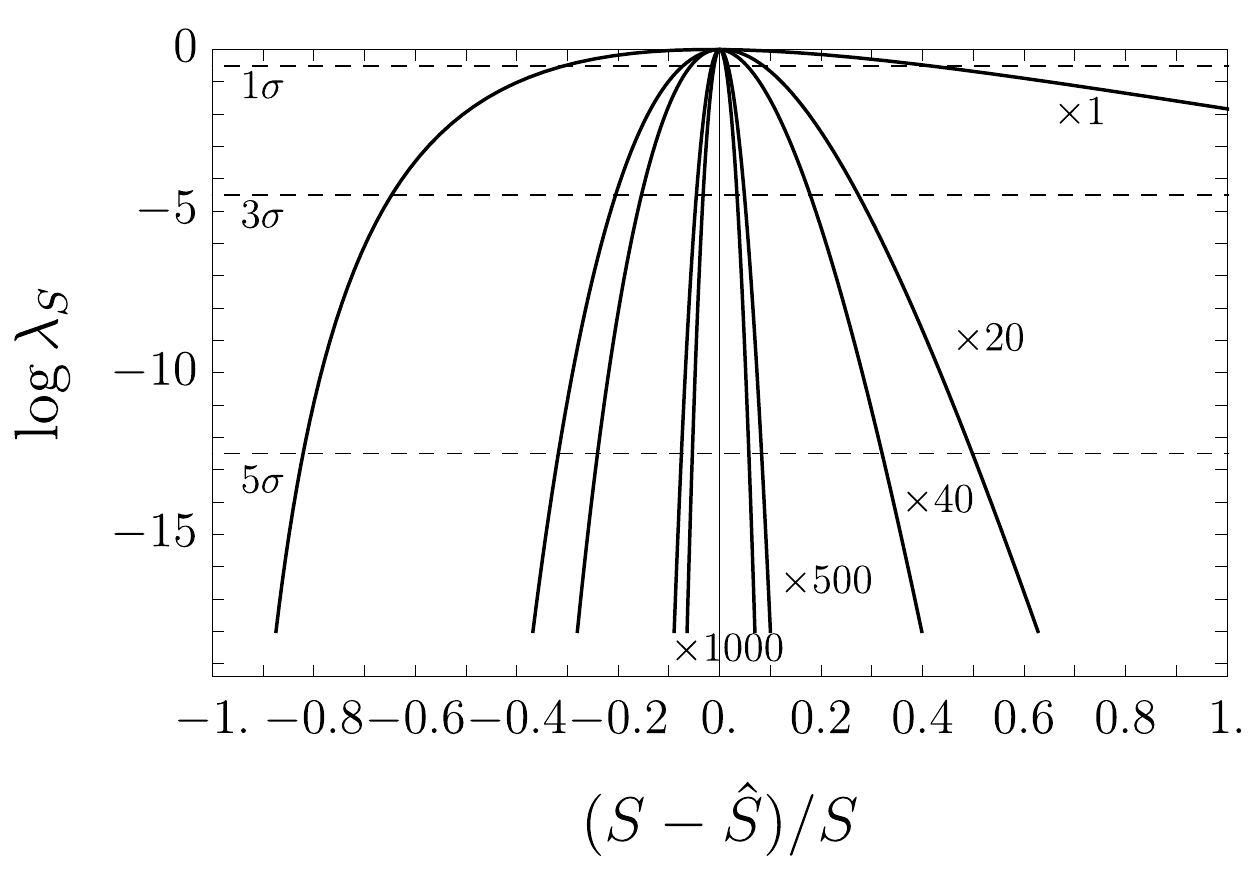}{0.9}
\caption{Profile likelihood ratio for $S$, of the IceCube data analysed in
  Fig.~\ref{fig:scale}, and the four simulated samples seen in
  Figs.~\ref{fig:gen2} and \ref{fig:supericecube}. Since different samples provide different estimates for $S$, the horizontal axis is rescaled to show all curves peaking at the same point, favoring visualisation.}\label{fig:profile}
\end{figure}

The likelihood maximizes for the pair of values
\begin{equation}\left\{\begin{array}{rcl}
S&=&1.16^{+0.51}_{-0.34}\, (1\sigma\,\mbox{C.L.}),\\[.2cm]
F&=&0.86^{+0.27}_{-0.22}\,
      (1\sigma\,\mbox{C.L.}).\end{array}\right.\label{eq:scaledr}\end{equation}
In Fig.~\ref{fig:scale} we show the confidence contours and the associated curves in the $F-S$ plane for each
event type that would produce the observed number of events of each
type. In Fig.~\ref{fig:profile} we show the profile likelihood ratio for $S$. Note that  the cross section is consistent
at the $1\sigma$ level with the value obtain from perturbative QCD
calculations guided by HERA data, and IceCube measurement~\cite{Aartsen:2017kpd,Bustamante:2017xuy}.  However, thus far the study is
statistics limited, with about 37\% uncertainty.  Note that because
we have combined various energy bins there
is a
dependence of the cross section with the flux normalization, but is
an almost negligible; see Appendix~\ref{app:2} for details.  Of course,  in a more general analysis considering an anisotropic flux of neutrinos and flavor ratios not equally distributed on Earth, additional free parameters need to be added to the likelihood analysis to account for the extra degrees of freedom.

\begin{figure}[tpb]
\postscript{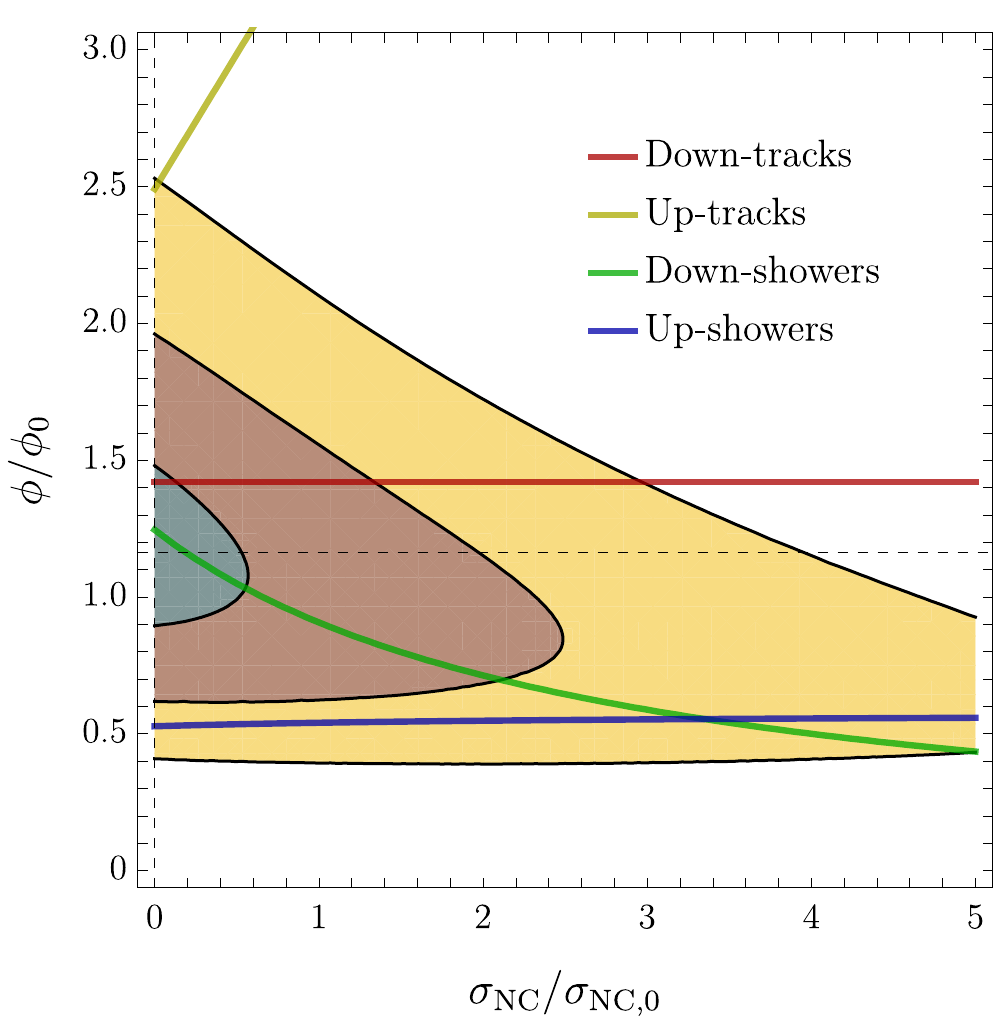}{0.9}
\caption{1, 3, and $5\sigma$ confidence contours in the $(F,S_{\rm SN})$  plane.
\label{fig:scale_nc}}\end{figure}

IceCube is also opening other doors to look for heavy new
physics. Even if the mean inelasticity measured by IceCube up to
$10^6~{\rm GeV}$ is in agreement with the SM
prediction~\cite{Aartsen:2018vez}, the energy dependence of the
neutrino-nucleon cross section~\cite{Bustamante:2017xuy} seems to
leave some room for {\it small} new physics contributions affecting
neutrino interactions both in~\cite{Ellis:2016dgb} and
beyond~\cite{Cornet:2001gy,Anchordoqui:2006wc,Barger:2013pla} the
SM. Note that all of these processes
would only increase the NC contribution to the neutrino-nucleon cross
section, thus modifying the (perturbative) SM prediction of the
$\sigma_{\rm CC}/\sigma_{\rm NC}$ ratio.

Next, in line with our stated plan, we duplicate our analysis but keeping the ratio
$\sigma_{\rm CC}/\sigma_{\rm NC}$ as a free parameter in the
likelihood function. The dichotomy between tracks
(which are only produced via CC interactions) and showers provides a direct test
of the $\sigma_{\rm CC}/\sigma_{\rm NC}$  ratio. Before proceeding we note that in the analysis carried out by the IceCube
Collaboration~\cite{Aartsen:2017kpd} only upward
going tracks are considered to keep the angular distribution of events with small
uncertainties. Because of this, their analysis sets a limit on the
charged-current neutrino nucleon cross section $\sigma_{\rm CC}$. In the analysis of~\cite{Bustamante:2017xuy} only the
shower-HESE data sample is considered, with full scrutiny of the
angular distribution of IceCube events. However, the ratio of the CC
and NC contributions is fixed to that expected in the perturbative SM, i.e. 
$\sigma_{\rm CC}/\sigma_{\rm NC} = 3$.

\begin{figure*}
\begin{minipage}[t]{0.49\textwidth}
\postscript{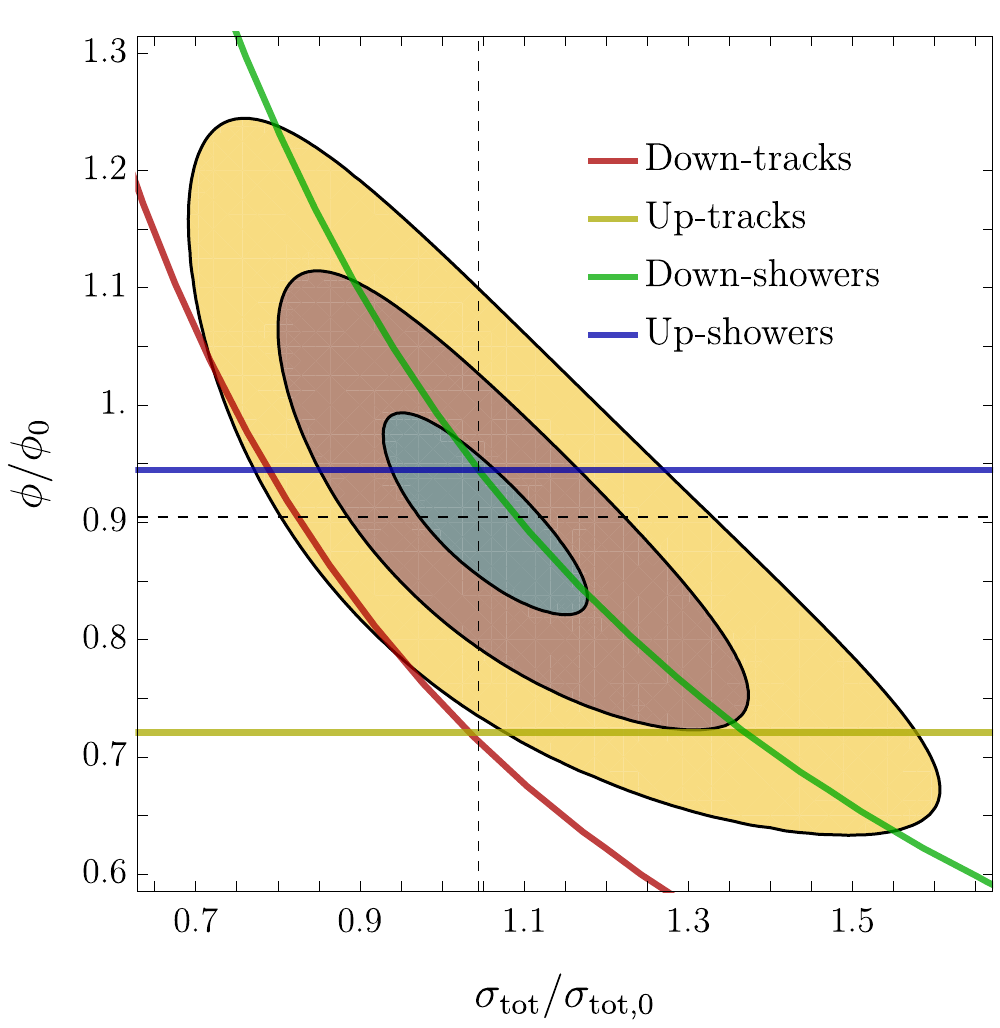}{0.9}
\end{minipage}
\begin{minipage}[t]{0.49\textwidth}
\postscript{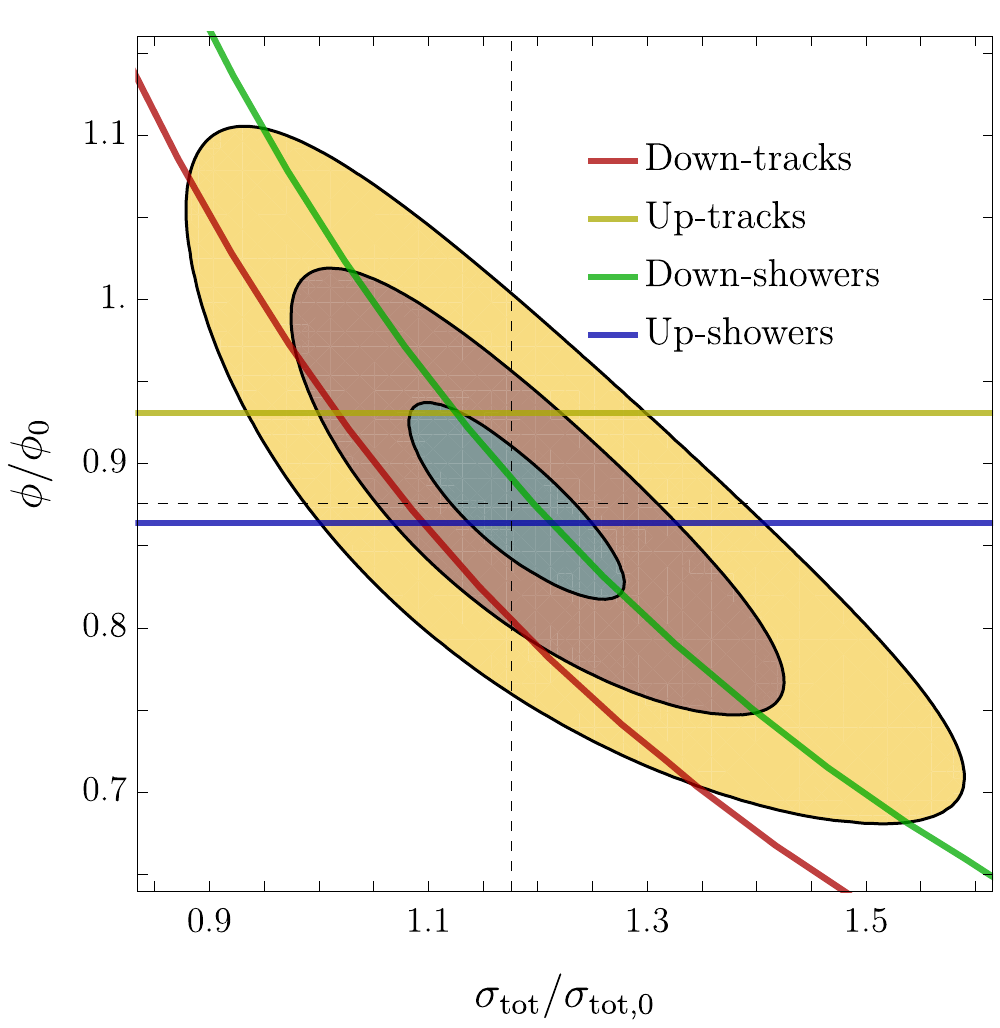}{0.9}
\end{minipage}
\caption{Projected determinations of neutrino fluxes and cross
  sections at $\sqrt{s} \sim 1~{\rm TeV}$ from future IceCube-Gen2
  data.  The 1, 3, and 5$\sigma$ confidence contours are based on
  simulated data for a $20\times$ (left) and $40 \times$ (right) the
  actual IceCube sample.}\label{fig:gen2}\end{figure*}

We begin by  writing the total neutrino-nucleon cross section as
$\sigma_{\rm tot}=\sigma_{\rm CC,0}+\sigma_{\rm NC}$. Instead of
considering the full scaling of the cross section $S$ as the parameter
of interest, we set out the analysis to constrain the ratio $S_{\rm NC}\equiv \sigma_{\rm NC}/\sigma_{\rm NC,0}$. Following a process similar to that used to obtain (\ref{eq:rates}) from (\ref{eq:rates0}), the expected numbers of down-going events are
\begin{eqnarray}
N_d^{\mathcal S} & = & F\left(\sum_{\alpha=e,\tau} \tilde N_d^{\rm
                       CC,\alpha}+S_{\rm
                       NC}\sum_{\alpha=e,\mu,\tau}\tilde N_d^{\rm
                       NC,\alpha}\right) \,, \nonumber \\
N_d^{\mathcal T} & = & F\,\tilde N_d^{\mathrm CC,\mu}.
\label{eq:ncdown}
\end{eqnarray}
Likewise, for up-going events, 
\begin{eqnarray}
N_u^{\mathcal S} & = & F \left(\sum_{\alpha=e,\tau} f_\alpha (S_{\rm
                       NC}) \tilde N_u^{\rm CC,\alpha}+S_{\mathrm
                       NC}\sum_{\alpha=e,\mu,\tau}f_\alpha(S_{\rm NC})
                       \tilde N_u^{\mathrm NC,\alpha}\right),
                       \nonumber \\
N_u^{\mathcal T} & = & F\,f_\mu(S_{\rm NC})\tilde N_u^{\mathrm CC,\mu},
\end{eqnarray}
where
\begin{equation}
f_\alpha(S_{\rm NC})\equiv\frac{1+r_\alpha}{Y_{\rm NC}S_{\rm NC}+r_\alpha},
\end{equation}
\begin{equation}
r_\alpha\equiv\frac{y^\alpha_{\rm CC,0}}{y^\alpha_{\rm
    NC,0}}\frac{\sigma_{\rm CC,0}}{\sigma_{\rm NC,0}} \,,
\end{equation}
and $Y^\alpha_{\rm NC}\equiv y^\alpha_{\rm NC}/y^\alpha_{\rm NC,0}$,
and where we have assumed that the average inelasticities for $\rm CC$
interactions remain unchanged from those of the SM. For the SM values, and assuming flavour independent inelasticities, we can approximate
 $r_\alpha\approx 8 \equiv r$. In such case, the up-going expected event numbers are simplified to 
\begin{eqnarray}
N_u^{\mathcal S} & = &F\frac{1+r}{Y_{\rm NC}S_{\rm NC}+r}
  \left(\sum_{\alpha=e,\tau}\tilde N_u^{\rm CC,\alpha}+S_{\mathrm
  NC}\sum_{\alpha=e,\mu,\tau}\tilde N_u^{\mathrm NC,\alpha}\right),
  \nonumber \\
N_u^{\mathcal T} & = & F\frac{1+r}{Y_{\rm NC}S_{\rm NC}+r}\tilde
                       N_d^{\mathrm CC,\mu} \, .
\label{eq:ncup}
\end{eqnarray}

Data from the Large Hadron Collider (LHC) put severe constraints on
stringy and gravity contributions to the neutrino-nucleon scattering
cross
section~\cite{Aad:2013gma,Sirunyan:2018xlo,Aaboud:2016ewt,Sirunyan:2017anm,Sirunyan:2018xwt}. However,
non-perturbative SM processes, such as sphaleron transitions, remain
almost unconstrained by LHC data~\cite{Sirunyan:2018xwt}. By comparing
the 90 fermionic degrees of freedom in the SM with the 6 degrees of
freedom in the neutrino sector contributing to missing energy
$/\!\!\!\! E_T$, we take $y_{\rm NC} \simeq 0.95$ and so using
$y_{\rm NC,0} \simeq 0.3$ we have $Y_{\rm NC} \simeq 3$. This
particular choice of $y_{\rm NC}$ is also valid for excitations of the
string and quantum black hole production in scenarios with large
extra-dimensions.\footnote{To first approximation, the black hole can
  be treated as a point-radiator that emits mostly $s$-waves. This
  indicates that it decays equally to a particle on the brane and in
  the bulk, since it is only sensitive to the radial coordinate and
  does not make use of the extra angular modes available in the
  bulk. Since there are many more particles on our brane than in the
  bulk, this has the crucial consequence that the black hole decays
  visibly to SM particles~\cite{Emparan:2000rs,Dimopoulos:2001hw}.}
However, this is not the case for exchange of Kaluza-Klein gravitons
in the large extra-dimension brane-world, where the transferred energy
fraction is only around
0.1~\cite{Kachelriess:2000cb,Anchordoqui:2000uh}.

Maximizing the likelihood (\ref{eq:like}) for the parameters $\boldsymbol\theta=\{F,S_{\rm NC}\}$ using (\ref{eq:ncdown}) and (\ref{eq:ncup})  provides the values

\begin{equation}\left\{\begin{array}{rcl}
S_{\rm NC}&=& 0.00^{+0.27}_{-0.00}\, (1\sigma\,\mbox{C.L.}),\\[.2cm]
F&=&1.16^{+0.20}_{-0.18}\,
     (1\sigma\,\mbox{C.L.}).\end{array}\right.
\label{eq:scaledr12}
\end{equation}
In Fig.~\ref{fig:scale_nc} we show the confidence contours and the
associated curves in the $F - S_{\rm NC}$ plane for each event type
that would produce the observed number of events of each type. We can conclude that $S_{\rm NC}>1$ is excluded at $2\sigma$
level. 

In summary, we have used the
complete (${\cal S}$ + ${\cal T}$) HESE data sample to constrain the
rise of $\sigma_{\rm NC}$. Because the
data are scant and the arrival direction of shower events have large
uncertainties we have chosen to
integrate over the angular distribution. Note that the analysis
presented herein is complementary to those
reported in~\cite{Aartsen:2017kpd,Bustamante:2017xuy} as it test a
different region of the neutrino-nucleon cross section parameter
space. Indeed, the likelihood fit given in (\ref{eq:scaledr12}) provides
the first {\it unequivocal} constraint derived from IceCube data on
non-perturbative SM phenomena~\cite{Ellis:2016dgb},
low-mass-string-scale Regge excitations~\cite{Cornet:2001gy}, and
gravity effects~\cite{Anchordoqui:2001cg}.

\subsection{Looking ahead with IceCube-Gen2}
\label{sec:gen2}

\begin{figure}[tpb]
\postscript{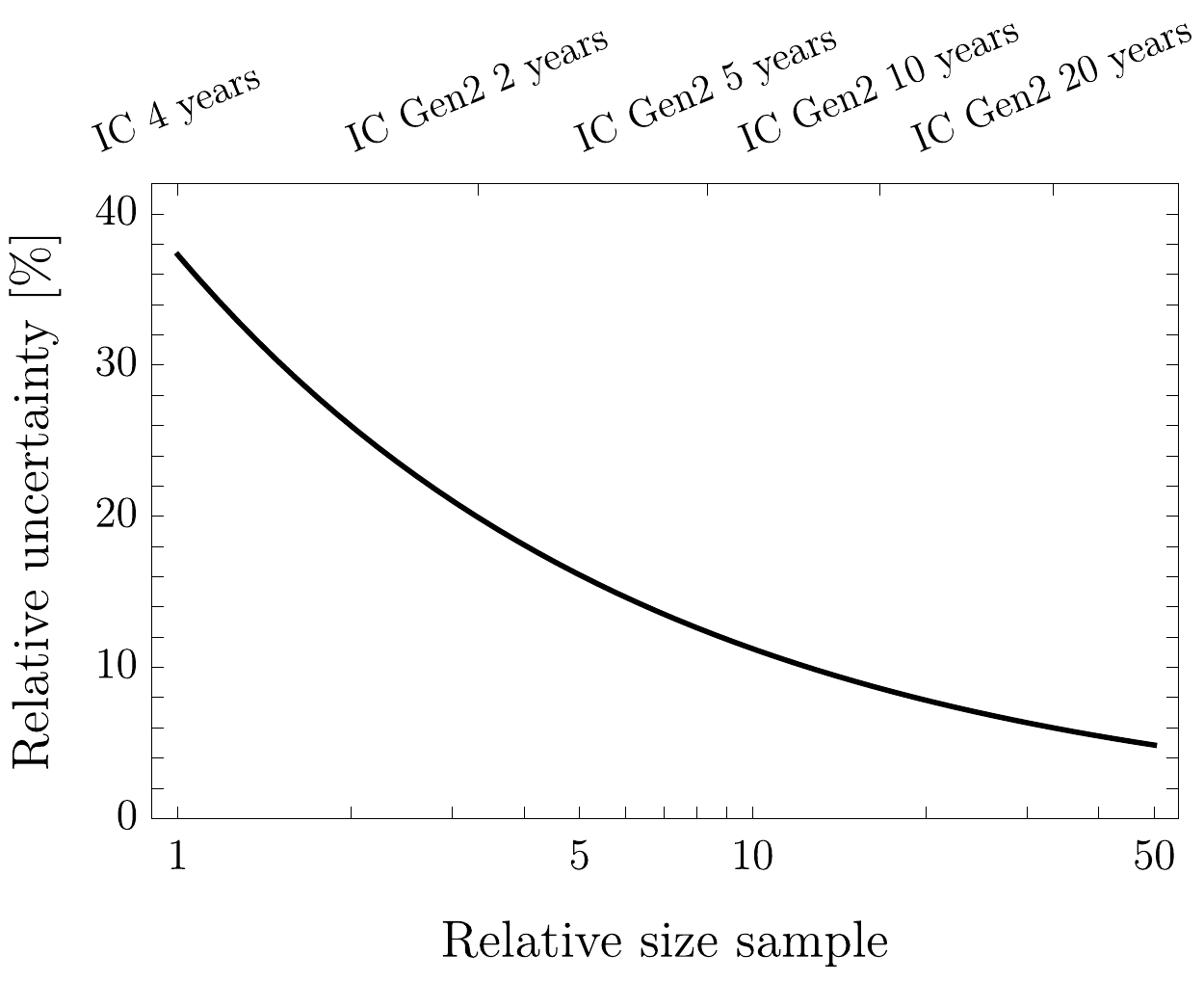}{0.9}
\caption{Evolution of the cross section precision measurement.}\label{fig:accuracy}\end{figure}

\begin{figure*}
\begin{minipage}[t]{0.49\textwidth}
\postscript{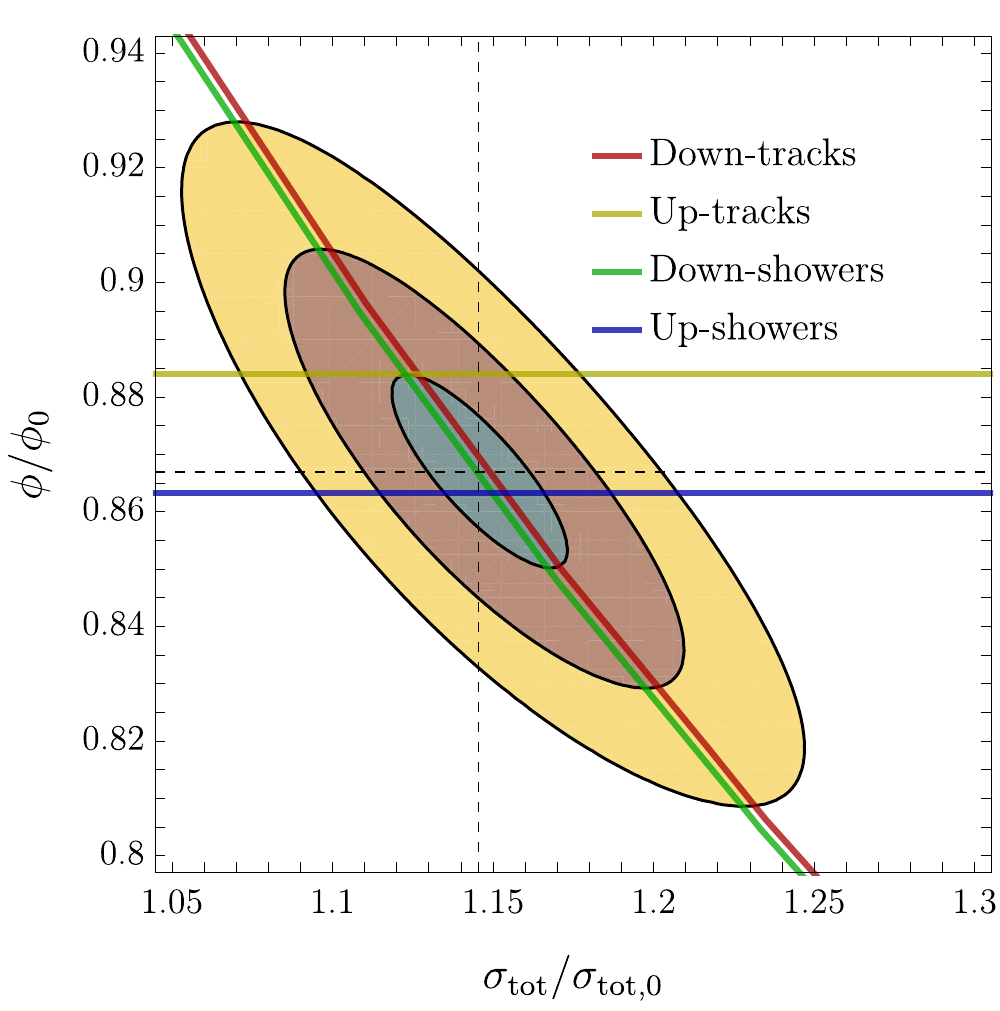}{0.9}
\end{minipage}
\begin{minipage}[t]{0.49\textwidth}
\postscript{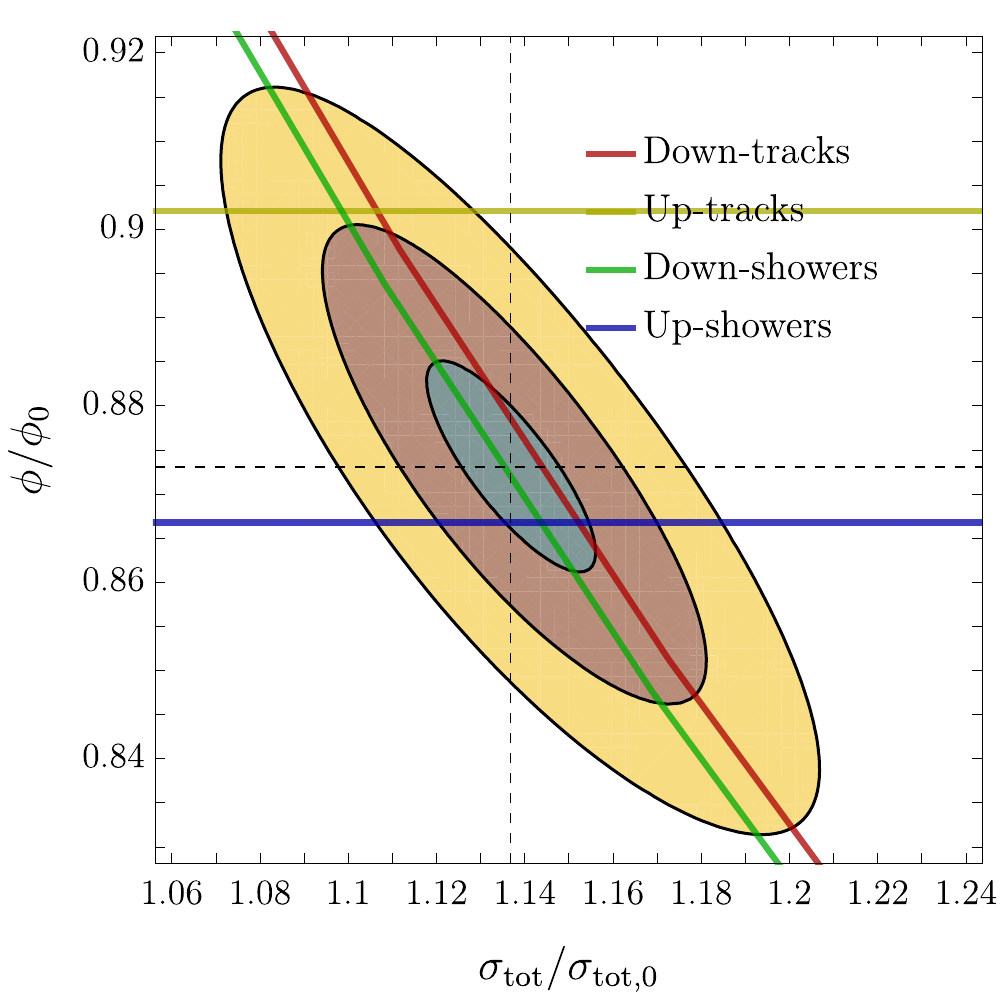}{0.9}
\end{minipage}
\caption{Confidence contours based in simulated data for a $500\times$
  (left) and $1000 \times$ (right) the actual sample.} \label{fig:supericecube}\end{figure*}

Design studies for the IceCube-Gen2 high-energy array are well
underway~\cite{Aartsen:2014njl}. They will result in an instrumented volume approaching
$10~{\rm km}^3$ and will lead to significantly larger neutrino
detection rates, across all neutrino flavor and detection channels. A
rough estimate indicates about an order of magnitude increase in
exposure per year. The bigger instrumented volume will facilitate the
calorimetric detection of muon tracks, reducing significantly the
systematic uncertainty. The extension will reuse the very reliable
design of IceCube's digital optical modules and therefore it will
surely perform technologically at least at the level of IceCube.  A
conservative estimate of the sample size is then attainable by simply
scaling the instrumented volume.

To determine the sensitivity of IceCube-Gen2 to probe strong dynamics,
we generate random samples of events, $\bar N^\mathcal Z _{x,k}$,
following the distribution (\ref{eq:poiss}), with the parameters for a
scaled total cross section found in the IceCube data analysis, which
are summarized in (\ref{eq:scaledr}). To
accommodate the bigger instrumented volume we adopt a multiplying
factor on the IceCube data sample. In 10 years of observation
IceCube-Gen2 will collect about 500 neutrino events in the energy
range $0.1 \alt E_\nu/{\rm PeV} \alt 2$, and about 1000 events in 20
years. Thus we adopt 20 and 40 as the representative multiplicative
factors associated with these data samples. Using the high-energy and
high-statistics sample to be collected by IceCube-Gen2, we perform the
same likelihood analysis as with the real data. The results are shown
in Fig.~\ref{fig:gen2} for a sample of $20\times$ and $40\times$ the
actual sample size. The precision on the cross section determination
would be 7.9\% and 5.5\% for $\sim500$ and $\sim1000$ events, respectively. This
precision is comparable to that obtained in perturbative QCD
calculations guided by HERA data. Of course this will also require a comparable
reduction on the systematic uncertainties, otherwise any study would become systematics-limited. Detailed evolution of the
uncertainty with sample sizes is illustrated in
Fig.~\ref{fig:accuracy}.

We can also envisage an IceCube-like detector of $100~{\rm
  km}^3$, specifically designed to probe strong dynamics. In
Fig.~\ref{fig:supericecube} we entertain this possibility and show the
results of a likelihood analysis based on simulated data for a $500
\times$ and a $1000 \times$ the actual sample. The $1\sigma$ contour regions
could reach a precision of less than 2\% level.

Some of the technologies needed to enable the next generation neutrino
observatories are already in development. For example, the strings of
IceCube-Gen2 will use multi-PMT DOMs, providing better directionality
and more than double the photocathode area per module~\cite{Kappes:2016kbd,Ackermann:2017pja,Classen:2019tlb}.  The new
instrumentation will dramatically boost IceCube-Gen2 performance.
The strings will feature new calibration devices that would allow to
better model the optical properties of the ice, reducing systematic
uncertainties in the tau neutrino appearance measurement, and
improving reconstruction of the direction of high energy cascade
events. The reduction of systematics uncertainties in the arrival
direction of ${\cal S}$-events would play a pivotal role in the
accurate determination of the neutrino-nucleon cross
section. Strategies and new devices to greatly improve the angular
resolution of next-to-next-generation cosmic neutrino detectors are
also under
discussion. Any detailed discussion addressing the challenges to be
faced in the design of these facilities falls outside the
scope of this article, which has been planned as a  phenomenological
approach to  neutrino scattering on ice.

\section{Conclusions}

\label{sec4}

Motivated by IceCube observations we have re-examined a technique to
probe strong dynamics with neutrino telescopes in the Antarctic
ice. The strategy involves comparing the rate for up-going and
down-going neutrino events to disentangle effects from the unknown
flux and those from QCD dynamics. More concretely, we implemented the
standard statistical analysis, using two uncorrelated observables (up-
and down-going events), to determine the best fit {\it model}
parameters (flux and cross section) and the fluctuations around the
favored values. The hypotheses of the {\it model} being tested are:
{\it (i)}~an isotropic neutrino flux and {\it (ii)}~flavor ratios
democratically distributed on Earth, both consistent with IceCube
data~\cite{Mena:2014sja,Aartsen:2015ivb,Palomares-Ruiz:2015mka,Vincent:2016nut,Aartsen:2017ujz}. Current
results from IceCube already provide interesting constraints on the
flux cross-section parameter space. Using 6~yr of IceCube HESE data we
have obtained a measurement of the neutrino-nucleon cross section at
$\sqrt{s} \sim 1~{\rm TeV}$. We have shown that the
measured cross section is consistent within $1\sigma$ with
perturbative QCD calculations constrained by HERA measurements, and
also with the recent IceCube
measurement~\cite{Aartsen:2017kpd,Bustamante:2017xuy}. Note that with
current statistics in the HESE data-sample our measurement has a 37\%
uncertainty, compared to the measured cross section with 35\%
uncertainty reported by the IceCube Collaboration using a larger data
sample, and the 5\% error of the SM experimentally constrained
calculation using HERA data. In a separate study we have also
constrained contributions from non-perturbative processes to the
neutrino-nucleon cross section. We have shown that contributions to
the NC interaction at $\sqrt{s} \sim 1~{\rm TeV}$ from electroweak sphaleron transitions are excluded at the $2\sigma$ level.

However, the most important result of our study is the investigation
on the potential of future neutrino-detection facilities for measuring
the neutrino-nucleon cross section. Using the energy and angular
distributions observed by the IceCube neutrino detector, we have
demonstrated that in the near future IceCube-Gen2 will carry striking
improvements to determine both astrophysical neutrino fluxes and cross
section.  In particular, we have shown that the high-energy
high-statistics data sample to be collected by this facility will
fetch a direct measurement of the neutrino-nucleon cross section at
$\sqrt{s} \sim 1~{\rm TeV}$, with a precision of about a $6\%$, that
is comparable to perturbative QCD informed by HERA data. We have also
shown that a $100~{\rm km}^3$ detector would reach the unprecedented
precision of less than a 2\% level.

\acknowledgments{We thank Subir
  Sarkar for permission to reproduce Fig.~\ref{fig:pQCD}. L.A.A. and
  J.F.S. are supported by U.S. National Science Foundation (NSF Grant
  PHY-1620661) and by the National Aeronautics and Space
  Administration (NASA 80NSSC18K0464). C.G.C. is supported by ANPCyT. Any opinions, findings, and
  conclusions or recommendations expressed in this material are those
  of the authors and do not necessarily reflect the views of NSF or
  NASA.}

\appendix

\section{Up-going event rate}
\label{app1}

The probability for a neutrino with incident angle $\vartheta$  measured from the
horizon and azimuth angle $\varphi$ to survive for a
distance $x$ along a chord through the Earth is given by
\begin{equation}
  P_{\rm survival} (x) = e^{-x/\lambda_a} \,,
  \end{equation}
where $\lambda_a = (\sigma_a \rho_\oplus N_A)^{-1}$ is the neutrino attenuation
length, with $N_A = 6.022 \times
10^{23}~{\rm g}^{-1}$ and $\rho_\oplus$ the Earth's density, and where
$ \sigma_a = \sum_i \sigma_i y_i$ is the attenuation cross section,
defined as the total cross section weighted by the inelasticity $y_i$,
with $i \in \{{\rm CC}, {\rm NC} \}$~\cite{Feng:2001ue}. The probability for neutrino
interaction producing an observable signal (either via a CC or a NC
process) in the interval $(x, x+dx)$ is 
\begin{equation}
P_{\rm signal} (x) =   \frac{dx}{\lambda_i} \,,
\end{equation}
where $\lambda_i =
(\sigma_i \rho_\oplus N_A)^{-1}$ is the
neutrino interaction length. The neutrino traverses a chord length $l = 2 R_\oplus \sin \vartheta$, with $R_\oplus$ the
Earth's radius. Note that $\lambda_a$ limits the maximum chord length,
and therefore the solid angle over which neutrinos can be observed scales
as $\Omega = 2\pi \sin \vartheta = \pi
\lambda_a/R_\oplus$~\cite{Hussain:2006wg}. For a given neutrino flux $\phi$,
the rate of up-going events at IceCube is then estimated to be
\begin{equation}
\Gamma_{\rm up}  = \phi \ A_\nu {\frac{d\varphi}{d\Omega}}
\int_{\ell}^{2R_\oplus} 2\pi\frac{dl}{2R_\oplus} \int_{l -\ell}^{l}
e^{-x/\lambda_a}\frac{dx}{\lambda_i} ,
\end{equation}
which on integration yields
\begin{equation}
\Gamma_{\rm up} = \phi \ \pi A_\nu {\frac{d\varphi}{d\Omega}}\frac{\lambda _a^{2}}{R_\oplus \lambda_i}
\left(1-e^{-{\ell}/{\lambda_a}} \right) \left(1-e^{-(2R_\oplus-\ell)/\lambda_a} \right) \,,
\label{eq:app4}
\end{equation}
where $A_\nu$ is the area of the detector projected against the neutrino
direction and $\ell$ is the portion of the neutrino path to which the
detector is sensitive~\cite{Kusenko:2001gj}. Note that the effective volume over which an
interaction may be detected is $V_{\rm eff} = A_\nu \ell$, where 
$\ell$ depends on: {\it (i)}~the lepton stopping ($dE/dx$) or decay,
{\it (ii)}~the chord length to the surface, and {\it (iii)}~the detector size.
For $2R_\oplus \gg \lambda _a \gg \ell$, (\ref{eq:app4}) simplifies to
\begin{equation}
\Gamma_{\rm up} \simeq  \phi \ \pi A_\nu {\frac{d\varphi}{d\Omega}}\frac{\ell}{R_\oplus}
\frac{\lambda_a}{\lambda_i} \propto \phi \frac{\sigma_i}{\sigma_a} \ . 
\end{equation}
For completness, we note that to calculate the event rate for a surface detector (e.g. the Pierre Auger Observatory), we must include an additional factor of 
$\sin\vartheta = l/(2R_\oplus)$ in the $dl$ integral 
to project out the 
normal component of the lepton flux emerging from the Earth, and so 
the rate of Earth-skimming neutrinos scales as $\phi \ \sigma_i/\sigma_a^2$, as shown in~\cite{Anchordoqui:2001cg}.

\section{Dependence on the neutrino spectrum}
\label{app:2}

As we have pointed out Sec.~\ref{sec2}, for a given bin of energy, we
can constrain neutrino interactions without assuming particular
neutrino fluxes or cross sections. However, because of the limited data sample
we have combined the results of various energy bins. This introduces a
dependence of the neutrino-nucleon cross section with the shape of the neutrino
spectrum.

\begin{figure}
\postscript{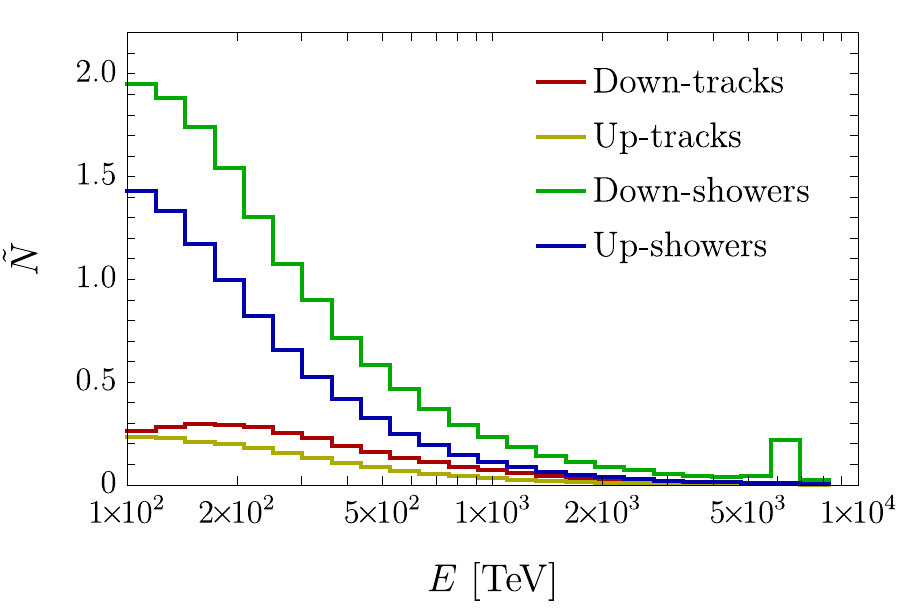}{0.99}
\caption{Reference number of events from (\ref{eq:reference}) with
  normalization given by (\ref{eq:fluxapp}).}\label{fig:app1}\end{figure}

\begin{figure}[tpb]
\postscript{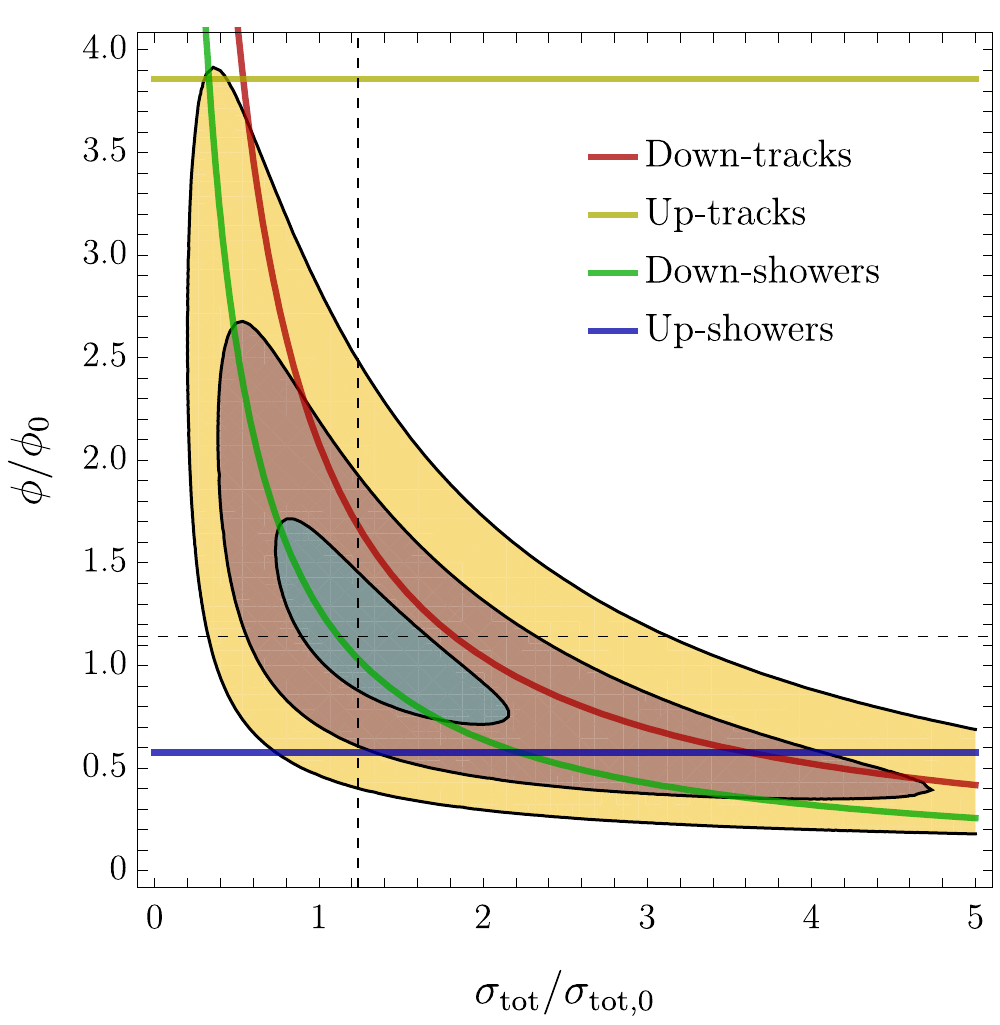}{0.9}
\caption{1, 3, and $5\sigma$ confidence contours for $(F,S)$ for
  scaled total cross section $\sigma_{\rm tot}$ and flux $\phi$ with
  respect to their reference values $\sigma_{{\rm tot},0}$ and $\phi_0$
  as given by (\ref{eq:fluxapp}).}\label{fig:app2}\end{figure}

To estimate the uncertainty associated with the spectral shape,  in what follows we
duplicate the procedure of Sec.~\ref{sec:3a}, but with a flux normalization given
by the most recent fit to the spectrum
by the IceCube Collaboration,
\begin{equation}
\phi_0 = N_{\phi} \
\left(\frac{E_\nu}{100~{\rm TeV}} \right)^{-\gamma}  \times 10^{-18}~({\rm
  GeV \ cm^{2} \ s \ sr)^{-1}} \,,
\label{eq:fluxapp}
\end{equation}
where $N_\phi = 6.45^{+1.46}_{-0.46}$ and \mbox{$\gamma =
  2.89^{+0.20}_{-0.19}$~\cite{Schneider:2019ayi}.} The values of the
expected number of events considering the central values of the flux
given in (\ref{eq:fluxapp}) are
shown in Fig.~\ref{fig:app1}. Table~\ref{tab:app} contains the expected number of events in each one of the four categories compared to the observed ones.

For the ratios given in Table~\ref{tab:app}, the likelihood maximizes for the pair of values
\begin{equation}\left\{\begin{array}{rcl}
S&=&1.24^{+0.54}_{-0.36}\, (1\sigma\,\mbox{C.L.}),\\[.2cm]
F&=&1.14^{+0.36}_{-0.30}\,
      (1\sigma\,\mbox{C.L.}).\end{array}\right.\end{equation}
In Fig.~\ref{fig:app2} we show the confidence contours and the associated curves in the $F-S$ plane for each
event type that would produce the observed number of events of each
type.

\begin{table}[h]
\caption{Observed/expected number of events in each category.}
\begin{tabular}{c|cc}
\hline
\hline
~~~~~Event direction~~~~~			& ~~~~~~~Shower~~~~~~~	& ~~~~~~~Track~~~~~~~	\\\hline
Down-going	&$18/14.1$		&$6/2.9$		\\\hline
Up-going		&$5/8.7$		&$7/1.8$
\\
\hline
\hline
\end{tabular}\label{tab:app}\end{table}

Note that by considering the steeper spectrum given in (\ref{eq:fluxapp}) the error in
the cross section slightly improves from 37\% to 36\%. We conclude
that the determination of the neutrino-nucleon cross section carried out
in Sec.~\ref{sec:3a} is robust.

\end{document}